# DESIGN OF A MULTIMODAL DEVICE TO IMPROVE WELL-BEING OF AUTISTIC WORKERS INTERACTING WITH COLLABORATIVE ROBOTS

**Carla Dei**
*Scientific Institute IRCCS Eugenio Medea (Italy)*
carla.dei@lanostrafamiglia.it

**Matteo Meregalli Falerni**
*National Research Council of Italy, STIIMA-CNR (Italy)*
matteo.meregallifalerni@stiima.cnr.it

**Matteo Lavit Nicora**
*National Research Council of Italy, STIIMA-CNR (Italy)*
matteo.lavit@stiima.cnr.it

**Mattia Chiappini**
*Scientific Institute IRCCS Eugenio Medea (Italy)*
mattia.chiappini@lanostrafamiglia.it

**Matteo Malosio**
*National Research Council of Italy, STIIMA-CNR (Italy)*
matteo.malosio@stiima.cnr.it

**Fabio Alexander Storm**
*Scientific Institute IRCCS Eugenio Medea (Italy)*
fabio.storm@lanostrafamiglia.it

## ABSTRACT

Several studies report that people diagnosed with autism spectrum disorder (ASD) have low employment rates and major difficulties in maintaining occupation (Hendricks, 2010). Adopting smart and interactive technology, promoted by Industry 4.0, can be a valid solution to improve social inclusion in the workplace. The present work describes the design and development process of the multimodal device (A)MICO (acronym for "A Multimodal device to improve inclusive Interaction between Cobot and Operator"), aimed to improve the user experience of ASD workers interacting with collaborative robots (cobots) in production lines. (A)MICO proposes a new intuitive mode of communication in which information about the cobot activity is transferred through acoustic and visual feedback. Lights and sounds are designed to make the user aware of the cobot's activity and gain control of the situation. To define the optimal feedback configuration, a co-design process has been performed involving five users with high functioning autism (HFA). The inclusion of users with HFA in the design process allows to analyse the system from different perspectives and increase the acceptance of the final product (Francis et al., 2009). In addition, (A)MICO is designed to be integrated into a pre-existing cobot system and its feedback can be customized making job tasks more human-friendly and accessible to all. To achieve this purpose, Design for All principles were taken into consideration since the preliminary stages of the study, in order to develop a device that could represent a valuable aid in improving the well-being and productivity for both neurotypical and neurodiverse users.

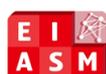

## KEYWORDS

Collaborative robots, autism spectrum disorder, work inclusion

## INTRODUCTION

Autism spectrum disorders (ASD) is a neurodevelopmental disorder characterized by repetitive behaviours and several deficits in social interaction (American Psychiatric Association, 2013; World Health Organization, 2013). Although several research studies highlight the presence of real talents among individuals with autism, there is still a focus on their impairments rather than on their strengths such as those that can be exploited in jobs (Johnson, 2022). Work experience of people with ASD can be made more complex by noisy environments and unpredictable social relationships (Lorenz et al., 2016). The lack of support, also caused by the overall organization of resources and environmental factors such as stigma, contributes to make more difficult finding a job for people with ASD (Unger, 2002). This social condition is subject to a lack of research focusing on employment of ASD workers (Johnson et al., 2020), which could in turn provide tools to make the workplace more comfortable for them. However, digital transformation brought by Industry 4.0 helps to provide a favourable context for work inclusion (Mark et al., 2019). The term "Industry 4.0" refers to the Fourth Industrial Revolution which is characterized by a progressive digitalization of the manufacturing process due to the new developed technologies. The fear to be replaced by the machine is reduced by the new human-centered work organization, which is also a core element of the concept of Industry 5.0: the operator becomes the centre of the production as a precious source (Rauch et al., 2020) and the machines are under his control. This new anthropocentric organisation led to a change in the criteria for recruiting workers and in the general work organization. In the context of Industry 4.0, the collaborative robots, or cobots, start to be adopted in the production lines. Cobots are machines that work side by side to the worker, implying the need to create new interaction modes (Weiss et al., 2021; Ávila-Gutiérrez et al., 2021). In human-robot interaction, the modality of communication has a big value also for facilitating the humanization of the cobot (Sciutti et al., 2018). This aspect is especially important when designing robotic systems that are more accessible for users who struggle with social communication. Several studies have investigated and demonstrated how, in this context, technology can be an effective support to facilitate inclusion in the workplace, promoting appropriate and customizable tools for people with ASD (Mpofu et al., 2019). The present work describes the design development of an integrative device, called (A)MICO, to improve and simplify the interaction between cobots and people with ASD involved in production lines. (A)MICO, through visual and acoustic feedback, aims to make explicit the implicit information of the system, in order to make the user more aware of the cobot on-going and future activities and therefore gain control over the situation. The biggest challenge has been evaluating what kind of information transferring and which is the best modality to do it without generating a negative impact on the ASD workers. In the creation and development process, the fundamental principles of "design for all" has been adopted to achieve a device that is a real help for all workers.

## METHODS

**Study Setting**

The present study was developed within the research activities promoted by the European project Mindbot (https://www.mindbot.eu/). The overall aim of Mindbot is to design human-robot interaction promoting mental health to facilitate active and positive job experiences in production lines using collaborative robots. (A)MICO was developed in the design and implementation phase of new technologies for the cobot platform. The initial phase of the project included also the organization of a laboratory setting, similar to an industrial one, where some participants, both neurotypical and high functioning ASD individuals, were involved in a collaborative task with a cobot performing a pick and place task. Participants were required to work with the cobot for three and a half hours for five consecutive days, to make their experience closer to that of operators in a real industrial scenario.

**Participants and Recruitment**

In order to investigate the needs and critical aspects of ASD users interacting with the cobot, five participants with ASD (1 female and 4 males) were involved in the preliminary design process of creating visual and acoustic feedback of (A)MICO. Each participant was individually involved in short interviews with one of the researchers. To enable the participant to get a complete overview of the topic, the interviews were usually organised after the last working day with the cobot.

**Co-design Process**

Co-design is a creative cooperation during design processes: the topic is approached from different perspectives and, including final user in the creative process, innovative design solution are achieved (Steen et al., 2011). In the present study, first of all, participant were asked for their opinion on working with the cobot, identifying strengths and weaknesses of the machine. Referring to the pick & place task performed with the cobot during the experimental trials, three situations were identified by the researchers as a potential source of stress: "the cobot stops because there is an error that requires external technical intervention", "the cobot stops because pieces are not placed correctly on the worktable" and "cobot confirms that the pieces have been placed rightly placed on the table". Indeed, working with complex technology sometimes generates in the worker a sense of incapability because of the feeling of not having complete control of the system. This situation had a negative impact on the emotional state of the person who begins to hesitate also in front of minor problems (Pollak et al., 2020). The importance of having additional feedback on the general progress of the work, emerged during the interviews with ASD users. After this step, for each situation mentioned above, some combinations of visual and acoustic feedback were displayed to the users by the researcher through an interactive questionnaire on a tablet (Fig.1). The colour code hypothesis was proposed considering its great impact on human-robot interaction, because it is the most intuitive way to transfer information about the activity of the cobot to the worker (Baraka et al., 2016). A quick benchmarking research has revealed that green, yellow and red are the most adopted colours in light signal systems integrated in cobots platform. In the present study, participants agreed on the choice of green and red but prefer to have orange instead of yellow due to the emission quality of led stripes integrated in (A)MICO and the environmental lighting.

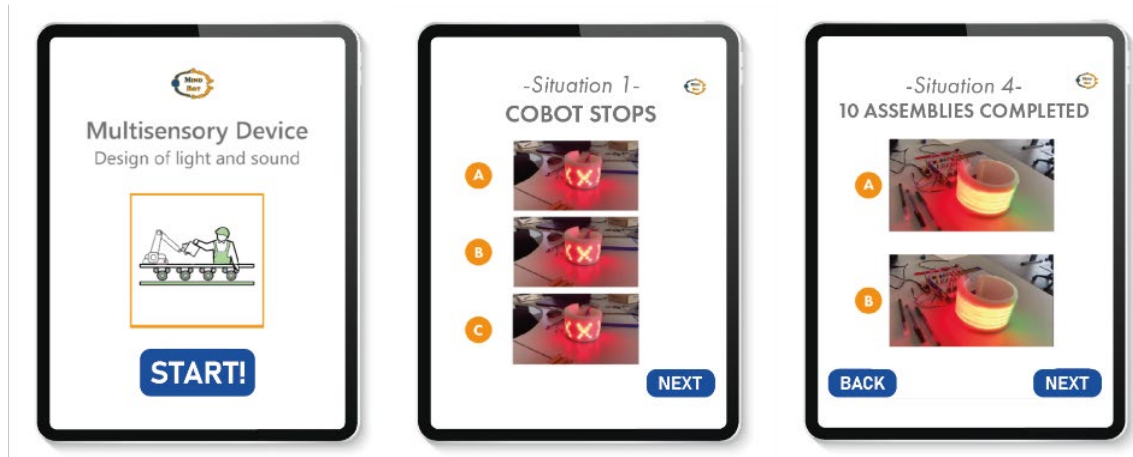

*Fig.1: Screens from the interactive questionnaire adopted to present various combinations of visual and acoustic feedback.*

# PRELIMINARY RESULTS

## Hardware & Product Design

The first prototype of (A)MICO was composed of a ring and a base. The ring was placed directly on the cobot and it had a flexible structure in order to be suitable for cobots of every size. Moreover, it was designed to be portable, without the need for power cables, using battery. On the inner part of the ring the stripes were fixed and placed as a matrix to show light signals in the form of a graphic pattern. The base, placed on the worktable, mainly served as station for recharging ring and as speaker's enclosure for emitting sound feedback. This kind of configuration, however, presented some problems due to the ring's power short-term power supply, and the message's clarity of the lighting patterns emitted. Considering these aspects, a second prototype of (A)MICO was developed, which consists of a tower integrating a speaker for sound emission and LED stripes disposed in the same matrix pattern as previously described (Fig.2). The tower is designed to be placed on the worktable, without taking up too much space and without disturbing the operator performing his work task but remaining visible to him when it is transmitting the information. (A)MICO is connected to the cobot system (ROS) (Quigley et al., 2009) and programmed via Arduino nano, translating the activities of the cobot into a lighting and acoustic pattern in real time.

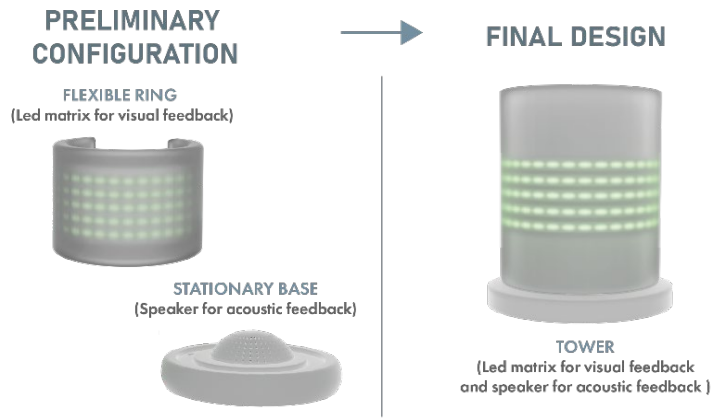

*Fig.2: Design development of (A)MICO. The preliminary configuration consisted of a ring and a base: in the final design, these two elements were combined into single one.*

**Feedback Design System**

The final choice of visual and acoustic feedbacks (Fig.3) is the results of the interviews performed with the five users with high functioning ASD who participated in the task with the cobot. Each graphic pattern has been designed to be always visible, from different points of view. The choice to transmit the information not only through a colour code but also through a graphic pattern enhances the meaning of the message and allows to be seen also by colour-impaired users. Regarding acoustic feedback, participants preferred soft and unobtrusive sounds. For the situation in which the cobot stops because of a system error, the feedback chosen is composed of red "X" signal blinking with a sound that recalls the idea of an alert. When the cobot pauses because it cannot identify the pieces, the feedback chosen is composed of blinking orange arrows pointing on the worktable, with a sound that recalls the idea of suspense. After the worker places correctly the pieces, the cobot restarts and all led blinking in green as a symbol of confirmation, together with a positive sound. The last kind of feedback was designed taking into consideration the need expressed by ASD users for frequent monitoring of work in progress. Every 10 assemblies completed, the LED matrix recreates the pattern of a rainbow and speakers emit a sound that represents a reward for the workflow continuity.

*Fig.3: Overview of the visual and acoustic feedback chosen by the ASD participants.*

# FUTURE PLANNED ACTIVITIES

At the current stage of the study, the second prototype of (A)MICO, characterized by its new 'tower' design, is ready to be tested in a controlled environment. If possible, future planned activities include involving ASD users who participated in the previous interviews to test the usability of the device, and to extend the tests also to other ASD users, not necessarily high functioning. Data collection will include practical cobot trial with and without (A)MICO; after the collaborative task, participants will be engaged in a semi-structured interview to have a feedback on the system usability and user experience. Considering the needs and issues related to ASD, ad-hoc questionnaires will also be adopted as support for the semi-structured interviews. The purpose of the interviews is to investigate ASD people's perception of (A)MICO.

# CONCLUSION

(A)MICO has been designed to provide the system's implicit information in a very intuitive way to make the users aware of the activity of the cobot and enhance their control of the situation, aiming to improve the general wellbeing and, consequently, the productivity. In addition, (A)MICO's configuration allows to customize the information's transmitted by LED light and speaker, making the device suitable for most cobot pre-existing systems and to satisfy the needs of a wide range of workers. The customization of the feedbacks makes (A)MICO a valuable aid to make workplace more accessible for everyone.

# AKNOWLEDGEMENTS

This research received funding from the European Union's Horizon 2020 research and innovation program under grant agreement No. 847926 (MindBot Project).